# Functional plasmonic capsules as colorimetric reversible pH-microsensors

Dr. Céline A. S. Burel[1], Alexandre Teolis[1], Dr. Ahmed Alsayed[1], Pr Christopher B. Murray[2], Dr. Bertrand Donnio[3], and Dr. Rémi Dreyfus[1]*

Dr. C. A. S. Burel, A. Teolis, Dr. A. Alsayed, Dr. R. Dreyfus
Complex Assemblies of Soft Matter Laboratory (COMPASS), UMI 3254, CNRS-Solvay-University of Pennsylvania, RIC, Bristol, PA 19007, United States
E-mail: rdreyfus@sas.upenn.edu

Dr. C.B. Murray
Dept of Chemistry and Materials Science, University of Pennsylvania
231 S. 34th Street, Philadelphia, PA 19104, USA

Dr. B. Donnio
Institut de Physique et Chimie des Matériaux de Strasbourg (IPCMS), UMR 7504, CNRS-Université de Strasbourg, 67034, Strasbourg, France

**ABSTRACT:** There is a crucial need for effective and easily dispersible colloidal microsensors able to detect local pH changes before irreversible damages caused by demineralization, corrosion, or biofilms occur. One class of such microsensors is based on molecular dyes encapsulated or dispersed either in polymer matrices or in liquid systems exhibiting different colors upon pH variations. They are efficient but often rely on sophisticated and costly syntheses, and present significant risks of leakage and photobleaching damages, which is detrimental for mainstream applications. Another approach consists in exploiting the distance-dependent plasmonic properties of metallic nanoparticles. Still, assembling nanoparticles into dispersible colloidal pH-sensitive sensors remains a challenge. Here, we show how to combine optically active plasmonic gold nanoparticles and pH-responsive thin shells into "plasmocapsules". Upon pH change, plasmocapsules swell or shrink. Concomitantly, the distance between the gold nanoparticles embedded in the polymeric matrix varies, resulting in an unambiguous color change. Billions of micron-size sensors can thus be easily fabricated. They are non-intrusive, reusable, and sense local pH changes. Each plasmocapsule is an independent reversible microsensor over a large pH range. Finally, we demonstrate their potential use for the detection of bacterial growth, thus proving that plasmocapsules are a new class of sensing materials.

1. Introduction

When corrosion occurs in a material, damages start from randomly located microscopic sites, where pH has already decreased.[1-3] When biofilms develop, bacteria change phenotypes and secrete substances that result in a local decrease in pH.[4-6] In the case of teeth, the more acidic environment causes local demineralization and is responsible for painful cavities.[7],[8] Although these three examples describe very different natural processes, they all carry one common feature: they describe heterogeneous processes, which start with local chemical changes at a microscopic level. Being able to track and detect these changes early open perspectives to prevent these processes and their subsequent adverse effects develop. Monitoring these local alterations requires the fabrication of smart microsensors that are cost-effective, easy to disperse, sense minute local pH changes rapidly, and able to translate these changes into a large, unambiguous signaling output. Optical signals have been used as output for microsensors because they are easy to read and visible to the naked eye without the need for a spectrophotometer.

So far, the dominant optical pH sensors able to reach some of the microsensors goals relies on the use of dyes such as fluorophores[9-15] or chromophores[11],[16],[17] which light absorption or fluorescence emission depends on the pH. By matching the adsorption peak of dyes with the emission wavelengths of nanoparticles (NPs) such as quantum dots, the fluorescence of the pH indicators can be greatly enhanced.[18],[19] Organic indicator dye molecules can be dispersed, adsorbed, entrapped, covalently bonded, or encapsulated in a support matrix.[5],[11],[13],[14],[20],[21] Though this type of sensors has proved very efficient, it usually requires careful monitoring of the dye self-aggregation, large concentrations, complex chemistry for their synthesis or their chemical attachment onto the polymeric backbone or a perfect control of the leaking for dyes encapsulation.[5],[11],[20],[21] Moreover, most dyes are sensitive to photobleaching, which limits their long-term use.[11] Furthermore, depending on the type of matrix and indicator, each pH sensor has a different measurable pH range, which usually goes only over a couple of pH units. Metallic NPs such as gold (Au) and silver (Ag) NPs absorb light in the visible due to their plasmon resonance.[22],[23] Additionally, plasmonic NPs extinction is extremely sensitive depending on NP composition, size, shape, orientation, local dielectric environment, and interparticle distance.[22-31] Furthermore, contrasting with organic dyes, plasmonic nanoparticles do not photobleach, can be easily synthesized in one single step process[32] and produced in large quantities. Based on their very interesting optical properties, their resistance to light and their easy synthesis, plasmonic NPs constitute a very good alternative to organic dyes. Therefore, a new generation of optical pH sensors made of plasmonic NPs and actuating polymers has been developed in recent years.[33],[34] The working principles of these sensors are based on the chemical and mechanical properties of the polymers used such as N-isopropyl-acrylamide-terminated hyperbranched polyglycerols[34] and carboxyl group-terminated poly(2-vinylpyridine)[33] and, the optical properties of the NPs used such as Au or Ag NPs. The combined assembly of optically active plasmonic NPs and pH-responsive polymers leads to an alternative class of optical pH sensors able to transform chemical potentials into a strong and easily detectable optical response. Although these systems are efficient, they do not provide a dispersible micron size sensor and are not suitable for real-life applications as their incorporation in different matrices strongly impact their optical response.

Here, we present an original alternative for the fabrication of highly sensitive and dispersible pH plasmonic microsensors. Briefly, these microsensors are obtained from Pickering emulsions made of Au NPs closely packed at an oil/ water emulsion interface.[35],[36] Then, the Au NPs are locked within a polyelectrolyte polymer shell at the interface giving the final shape to the Au NPs-polymer microcapsules' sensors, thereafter referred to as plasmocapsules. We demonstrate how these microsensors show a significant optical response to pH changes due to their shift in the plasmonic extinction maximum. As schematically shown in Figure 1, as the polymeric matrix swells or shrinks upon pH change, the distance between neighbored Au NPs, and thus their interparticular coupling, changes, resulting in a different optical response.[33] In our emulsion templated method, billions of microsensors are fabricated simultaneously, each plasmocapsule acting as a single micron-size sensor. These colloidal microsensors are easy to fabricate, dispersible, robust, non-intrusive, reusable, reversibly sense minute local changes over a very large pH range (7 pH units), and able to translate these changes rapidly into a large, unambiguous signaling output. As proof of concept, the use of these colloidal microsensors for the detection of acid-producing bacteria is demonstrated. We anticipate that our approach enables the development of a large body of similar hybrid microcapsule sensors for tracking and sensing other stimuli in spatially homogeneous and heterogeneous systems.

## 2. Principle of the Synthesis

The precise synthetic protocol is described in details in the Methods Section and relies on the following methodology: Au NPs functionalized with poly(diallyldimethylammonium-nitrate-co-1-)vinylpyrrolidone (PVP DADMAN, full structural formula available in Figure S1 of the Supporting Information), are synthesized by reduction of Au salt with ascorbic acid[32],[37]. The Au NPs obtained are polydispersed, the mean diameter d is d = 30 nm ± 11 nm and the PVP-DADMAN capping layer is about 1.5 nm (TEM images, histogram of size distribution, and extinction spectrum of the Au NPs can be found in Figures S2 of the Supplementary information). In this size range, the polydispersity does not affect the position of the plasmonic peak but results in a small broadening of the peak.[22],[37] Then, a solution of toluene containing a mixture of monomers of methyl methacrylate (MMA) and butyl acrylate (BA) (50:50 w/w), is emulsified in a continuous aqueous phase of the previously synthesized Au NPs. By setting the pH of the solution to 1.2, electrostatic barriers are removed, and the NPs adsorb at the surface of the emulsion due to energy minimization, forming a so-called Pickering emulsion of Au NPs[37] (Figure 2a). The red color of the emulsion at the beginning of the process indicates Au NPs adsorption (Figure 2b). As synthesis proceeds, color change from red to blue acts as a good visual indicator of the NPs close-packed arrangement at the emulsion interface. Indeed, only a tight and dense packing results in the coupling of the plasmonic resonances of neighbored Au NPs, giving a strong blue color to the emulsion[22],[38]. The polymerization of the encapsulated monomers occurs at the interface by raising the temperature above 60°C, giving rise to blue capsules (Figure 2c) made of an elastic shell in which gold nanoparticles are very closely embedded (Figures 2d). The plasmocapsules' spherical shells are observed by SEM. A high magnification image of the shell shows the dense packing of the nanoparticles (Figure 2e). The thickness of the shell is measured from the SEM images to be approximately 100 nm +/- 20 nm (see Figure S3 in supplementary information).

Mass spectroscopy is performed to quantify the amount of methyl methacrylate (MA), and butyl acrylate (BA) polymerized. Results show that essentially all of the two monomers are consumed (98.3% of MMA and 95.4% of BA, see Figure S4 in supplementary information). Upon addition of sodium hydroxide, about 24 mol% of the acrylate and carboxylic acid moieties present on the polymer shell are transformed into carboxylate ions (see the Supplementary information section for more details on the characterization of the plasmocapsules' polymer shell). The pH sensitivity of the polymer[39],[40] in which the Au NPs are embedded, makes the collective assembly of nanoparticles sensitive to any pH-changing process. As pH is increased, the presence of numerous free carboxylic groups on the surface confers charges to the shell, resulting in osmotic swelling of the shells. The whole construct made of Au NPs embedded in a pH sensitive crust can then be used as a pH-microsensor, each capsule acting as an independent sensor.

3.  **Experimental Results and Discussion**

The hypothesis that the synthesized microcapsules are de facto pH sensitive needed to be demonstrated first. To do so, we performed microfluidics experiments. We first introduced an acidic solution (pH ~ 1.2) containing blue microcapsules in a capillary, and then, a small droplet of a basic solution (pH ~ 13.8) was deposited gently at one end of the capillary. The droplet locally changed the pH, generating a front of increasing pH that propagated in the capillary. This front propagation could be easily followed by optical microscopy, the microcapsules changed colors as the pH solution was increased (Figure 3a, images 1-5, plasmocapsules in the capillary under different pH locally exhibit strong red and blue colors in the regions of high and low pH, respectively. The propagation of the pH front can be seen in movie 1 in the supplementary materials). At low pH, the polymeric backbone is uncharged, the nanoparticles are densely packed in the crust, and thus strongly coupled, and the microcapsules are blue. At higher pH, the polymer acquires charges (hydrolysis of acrylic acid –m- and acrylate ester –n and p- groups into carboxylate groups –m'-, see Figure S8 in supplementary information), the swelling crust increases the interparticle distance between the Au NPs, the plasmons are no more coupled and the microcapsules appear red. We also showed that this color change could even be monitored at the level of a single capsule (Figure 3b, images 1-10, and movie 2 in the supplementary materials).

To ensure that the change of color is solely due to the mechanical swelling of the polymeric crust embedding the Au NPs, we measured and plotted the color and size of the microcapsule as a function of time as the pH front propagation. It can be seen that as the propagation front moves with time over a microcapsule (Figure 4a), both the red intensity and the capsule diameter increase. When the red intensity was plotted as a function of capsule size (inset Figure 4a), a clear correlation between size and color could be found: a relative diameter change of about 10% is found sufficient to induce this color transition. To preclude the possibility that leaking or desorption of the Au NPs from the polymeric matrix might be responsible for the color change, we also check the reversibility of the color transition. We show that by changing the pH inside the microfluidics chamber back to acidic, the color of the microcapsules turns back to blue (Figure 4b). Moreover, the color of the capsules can be cycled, at least three times, from blue to red to blue, and so on, attesting of their flexibility and robustness (Figure 4c). This simple set of experiments evidences the pH-sensing potential of the microcapsules and how the coupling between the mechanical properties of the polymeric crust and the optical properties of the Au NPs allows us to obtain an easy read of local pH changes.

For the calibration of the colloidal microsensors, we performed a series of experiments in which the microcapsules were prepared and redispersed in different buffers of pH ranging from 2 to 12 (see the materials and methods section). They were first synthesized following the protocol described in the materials and methods section. Once they were formed, the capsules were redispersed in several solutions of different pH. The pH was measured using a microelectrode connected to a pH-meter. Direct observations of the microcapsules under the microscope confirmed that the color of the capsules was dependent on pH. Typical images of the plasmocapsules at five different pH values (i.e., pH = 2.4, 4.4, 7.4, 8.2, and 11.2) are showed in Figure 5. The color of the capsules is triggered by the change of pH in the surrounding suspension. To be more quantitative, for each pH sample, about 50 images were taken using a color camera, and the red, blue and green intensities were extracted from the camera channels and plotted as a function of pH (Figure 5a). A large red color transition in a narrow zone of pH, which extended from pH = 7.2 to pH = 8.2, was observed. The blue and green intensities remained quasi-constant at all pH values. The transition and the distinction between pH below 7.2 and pH above 8.2 are observable by the naked eye. In the pH range of 2 to 9, microcapsules color change due to pH increment as small as 0.5 pH unit were distinguished by eye under the microscope. As seen on the images of Figure 5 (a), a large range of colors is observed: blue, purple, pink, red, and strong red. The diffusion of hydronium ions over a 100 nm polymeric shell can be estimated by taking $10^{-10}$ m$^2$/s as typical value for the diffusion coefficient.[41],[42] The time it takes for a thin 100 nm pH-responsive polymer membrane to swell is then of the order of 0.1 ms: the membrane swelling of the plasmocapsules is almost instantaneous. This is consistent with the immediate change of color observed by eye under a microscope when a droplet of the base is added to an acidic droplet of plasmocapsules. This shows that the capsules can be indeed used to detect or monitor processes, which alter the pH of surrounding solutions in a large pH-range.

Finally, to demonstrate the applicability of these pH-sensitive plasmocapsules as efficient optical sensors, we used them to detect a natural process and replaced usual microelectrode measurements by simple in-situ straightforward optical measurements. As proof of concept, we used them to monitor pH change when a bacterial growth process occurred in a solution. When certain bacteria strains grow in a solution with nutrients, they can metabolize surrounding sugars, and degrade them into acids. The increasing concentration of acid induces a decrease in the pH of the medium. Measuring a pH change is thus one way to diagnose the presence of a living and growing strain in a contaminated solution. A strain of *Staphylococcus aureus* was grown in a nutrient broth containing 1 wt% D-(+)-glucose. The bacteria were incubated for two days at 35°C. The broth alone (no bacteria) was also incubated and the pH was controlled during the entire duration of the experiment. *S. aureus* colony forming units (CFU)/mL was determined at different times by serial dilution, and plating. Simultaneously, the pH was measured both in the broth alone (without bacteria) and in the broth containing bacteria (see Figure S9). Over time, the *S. aureus* grew from $10^3$ CFU/mL to $10^8$ CFU/mL, the exponential growth phase, and the stationary phase respectively occurring in about 10 and 28 h after incubation. Concomitantly, due to the fermentation of glucose by *S. aureus*, the pH was found to decrease from 8 to 4.5. The larger pH drops occurred during the exponential phase when bacterial growth was the most important, the pH stabilized at 4.5 once the bacteria reached the stationary phase. As a reference, the pH of the broth alone containing no bacteria remained constant with time, showing that the observed pH decrease is only due to the fermentation of glucose by *S. aureus*. The plasmocapsules were then dispersed in the culture media at a different time of the incubation, and the different colors were analyzed. As the bacteria grew in solution and the pH subsequently decreased, the microcapsules changed color from red at pH 8 to purple at pH 4.5. The Red, Blue and Green intensities of the microcapsules were plotted versus the concentration of *S. aureus* (see Figure 5b). We observed that as the population of the bacteria increased, the red intensity decreased, and the blue and green intensities remained constant. A color shift in the red intensity could be measured from $10^6$ bacteria in solution, setting the detection limit of this method. Therefore, the plasmocapsules can be used to detect the presence of acid-producing bacteria by sensing pH variations.

4. **Conclusion**

We have demonstrated the synthesis of pH-sensitive plasmonic microcapsules. They represent a new type of dispersible optical sensors made of a combination of a pH-sensitive polymeric shell, which swells or shrinks under pH changes, and embedded plasmonic nanoparticles, the optical response of which depends on the interparticle distance. Using regular colorimetric analysis, we have shown a direct correlation between the capsules color d the size of the plasmocapsules. We have calibrated the relation between the outside pH and color of the plasmocapsules. We have demonstrated that the optical response of the plasmocapsules is contrasted enough to detect different pHs, even though we use a path of synthesis that yields polydisperse nanoparticles. The plasmocapsules sensor allows for detection of pH in a relatively large range of 2 to 9. The plasmocapsules microsensors response time is of the order of 0.1ms, which is much smaller than competing organic dye technologies whose typical response time range from several tenths of seconds to a minute.[11] In the case where the microsensor would be a 10 microns particle of polyacrylic acid with Au NPs on its surface, the response time would be increased to about 1s. This validates the use of a thin shell of pH-responsive polymer instead of a large particle. We did not observe leaching of the NPs out of the polymeric shell, which was evidenced by the reversibility of the capsules over at least three cycles. Finally, we have shown that the plasmocapsules can be used to probe and follow the growth of bacteria, which change their environmental pH as they replicate. Because of all these characteristics, the pH-responsive plasmocapsules could be of interest to detect and monitor bacterial contamination in health care and the food industry.

The microsensors showed here were specially developed for pH measurement, however we believe that the effect and principle demonstrated here are general enough to be extended to other types of detectors. Swapping a pH-sensitive polymeric shell with, for instance, a fluoropolymer one, which, depending of the fluorination degree, would be able to selectively dissolve a wide range of gases ($H_2$, CO, $CO_2$, $NO_2$, $CH_4$) and vapors' solvents, would allow the fabrication of very specific gas sensors.[43] Alternatively, a polymeric shell sensitive to *e.g.* polynitroaromatic compounds would be of interest in the area of explosive sensing.[44] Using a temperature sensitive polymeric shell, which would swell/shrink under small temperature variations, could also be used as a temperature sensor in the cold chain to ensure fresh and frozen aliment preservation, chemical hazards limitation, or drugs stabilities. Therefore, the nature of the shell provides the type of detection to be done while the optical reading would still lie on the plasmonic properties of the embedded nanoparticles. As such we anticipate that these microcapsules will likely represent a new class of cheap non-intrusive microsensors able to transform chemical potentials or external forces into a large optical absorption shift, and easily detectable by the naked eye.

## 5. Experimental Section

**Materials:** Gold (III) chloride trihydrate (99.9%, Sigma-Aldrich), L-ascorbic acid (99%, Sigma-Aldrich), hydrochloric acid (ACS reagent, 37%, Sigma-Aldrich), toluene (Chromasolv Plus, 99.9%, Sigma-Aldrich), poly(diallyldimethyl-ammoniumnitrate-co-1-vinylpyrrolidone (PVP-DADMAN, Solvay®), Sodium Hydroxide (50/50 w/w Baker), methyl methacrylate (99%, Sigma-Aldrich), butyl acrylate (99%, Sigma-Aldrich), 2,2′-Azobis(2-methyl-propionitrile) (AIBN) (98%, Sigma-Aldrich), citric acid (ACS reagent, EMD Chemicals), sodium phosphate dibasic (ACS reagent, VWR), sodium carbonate dibasic (99.5%, ACS reagent, Sigma-Aldrich), sodium bicarbonate (99.7%, ACS reagent, Sigma-Aldrich), D-(+)-glucose (99.5%, Sigma Aldrich), sodium chloride BioXtra (99.5%, Sigma Aldrich), sodium hydroxide pellets (98%, Alfa Aesar), Nutrient Broth powder (NEOGEN Corporation), 9 mL

Nutrient Broth tubes (General Laboratory Products), Tryptic Soy Agar plates (USP G60, Hardy Diagnostics). The bacterium used in the studies was *Staphylococcus aureus subsp. aureus* (ATCC 6538). Ultrapure water (type 1, 18.2 MΩ.cm) was used for all culture media preparation and deionized water was used otherwise.

**Gold nanoparticles synthesis:** The Au NPs used in this procedure were synthesized by reduction of $HAuCl_4$ with ascorbic acid in presence of PVP-DADMAN. A solution of $5.7.10^{-6}$ M PVP-DADMAN and $2.510^{-3}$ M $HAuCl_4$ was brought to boil. Then, 12.5 mL of ascorbic acid (0.1 M) was added. The solution protected from light by aluminum foil was stirred for 1 h at 97°C. After synthesis, the Au NPs were left to rest for one day to remove the biggest nanoparticles. The rest of the dispersion was centrifuged and concentrated into a few milliliters' solution. MilliQ water was used throughout the entire synthesis process.

**Plasmocapsules synthesis:** An aqueous phase containing the Au NPs ($[Au^0]$ = 0.02 M) and HCl (20 μL) is mixed with a toluene phase (0.5 mL) of the dissolved acrylate monomers (MMA 2.2 M and BA 1.7 M) and AIBN (0.12 M). The two phases are strongly agitated by sonication with a Brandson 3210 ultrasonic bath at 30°C for 15 minutes, forming an oil-in-water emulsion. The formation of the polymer shell by polymerization of the monomer at the interface of the emulsion is carried out at 62°C for 2hrs.

**Plasmocapsules in buffer solutions:** buffer solutions were prepared by mixing the following four solutions in different ratios: Citric acid (0.75 M), $Na_2HPO_4$ (1.5 M), $Na_2CO_3$ (0.75 M) and $NaHCO_3$ (0.75 M). Buffer solutions at a specific pH were mixed with the dispersion of microcapsules in a 50/50 v/v ratio, and the pH was controlled afterward. Each time the final pH value was close (within 0.5 unit of pH) to the pH of the buffer solution. For high (> 11) and low (< 3) pH, NaOH and HCl were respectively used to adjust the pH of the plasmocapsules dispersion.

**Electron microscopy:** Scanning electron microscopy (SEM) was performed on JEOL 7500 HRSEM. The samples were prepared by air drying under ambient conditions by laying a drop of microcapsules dispersion in water on an electron microscopy science carbon-coated copper grid. Transmission electron microscopy (TEM) was carried out on a JEOL JEM-1400. The accelerating voltage was set at 120 kV.

**Optical characterizations:** Optical microscopy images were acquired on an Olympus microscope in bright field and transmission mode.

**pH measurements:** The pH of all solutions was measured with pH-meter (model 8015, VWR International) with a micro pH combination electrode (Z113425-1EA, Millipore Sigma).

**Bacteria cell culture and plate counting:** A fresh culture of *S. aureus* was prepared by inoculating 9 mL Nutrient Broth with a single colony from a Tryptic Soy Agar plate streaked for less than two weeks. Then, culture was incubated overnight at 35 °C under shaking conditions (250 rpm) in an incubating mini shaker (12620-942, VWR International). Bacteria were harvested by centrifugation for 10 min at 2000×g, washed twice using the medium studied, before resuspending in the medium studied. The concentration of *S. aureus* was determined by serial dilution with subsequent plating on Tryptic Soy Agar plates and measurement of colony forming units (CFUs) after 24 h at 35 °C.

**pH-bacteria growth study in planktonic conditions:** Two sterile culture media were prepared: a Nutrient Broth with 1 wt% D-(+)-glucose, 3 wt% sodium chloride; and a Nutrient Broth with just 3 wt% sodium chloride. Both solutions were adjusted to pH 8 with NaOH and filtered using 0.2 μm PES sterile syringe filters (VWR International). Bacteria were washed twice and resuspended in both solutions. Initial *S. aureus* concentration was adjusted by dilution at around $10^3$ CFU/mL. Then, bacteria were cultivated in closed vials on a MaxQ4000 shaker (model 4342, Thermo Scientific) for 48 h at 35 °C with shaking (100 rpm). *S. aureus* concentration was determined by plating and pH was measured. The microcapsules were concentrated and washed several times with aliquot of the bacteria solution at a different time of incubation until the pH of the final dispersion matches the pH of the bacteria solutions. Then the microcapsules were imaged by optical microscopy.

**Supporting Information**

Structural formula of PVP-DADMAN; SEM image of a plasmocapsule polymer shell; HPLC chromatograms obtained for plasmocapsules at pH 1.2 and at pH 13.8; characterizations of the plasmocapsules polymer shell with FTIR spectra, GC and GC-MS chromatograms; schematic representation of the color change of the plasmocapsules upon swelling; variation of the pH and color of the plasmocapsules as a function of the concentration of *S. aureus*; additional materials and methods. Supporting Information is available from the Wiley Online Library or from the author.

6. **Acknowledgments**

The authors thank the ANRT, GIE AIFOR, CNRS, and Solvay for financial support. The authors acknowledge the support of the French National Agency of Research (ANR) to the project REACT through the grant ANR15-PIRE-0001-06. SEM imaging was performed in facilities supported by the NSF MRSEC program under award No DMR-1720530. The authors also thank D. Bendejacq, R. Giordanengo, L. Gage, D. Radtke, C. Anderson, J. Hutchinson, and A.J. Yodh for fruitful discussions.

Received: ((will be filled in by the editorial staff))
Revised: ((will be filled in by the editorial staff))
Published online: ((will be filled in by the editorial staff))

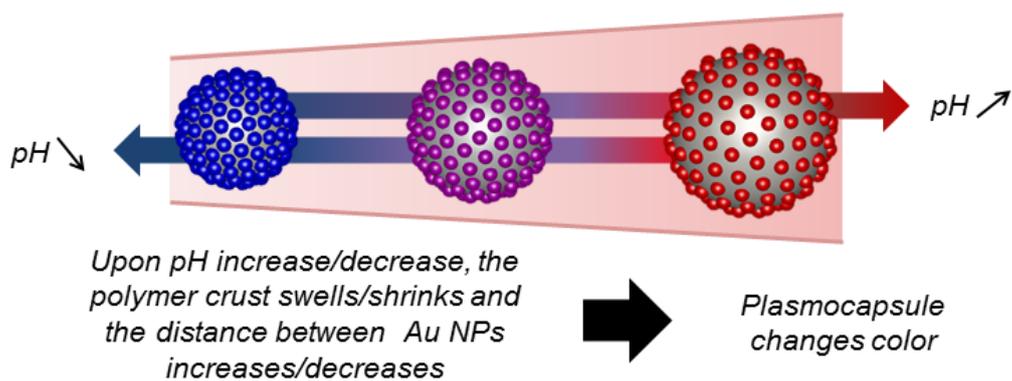

**Figure 1**: Schematic representation of the plasmonic response of the designed plasmocapsule microsensor as a function of pH change. As pH increases the plasmocapsule swells, and change color.

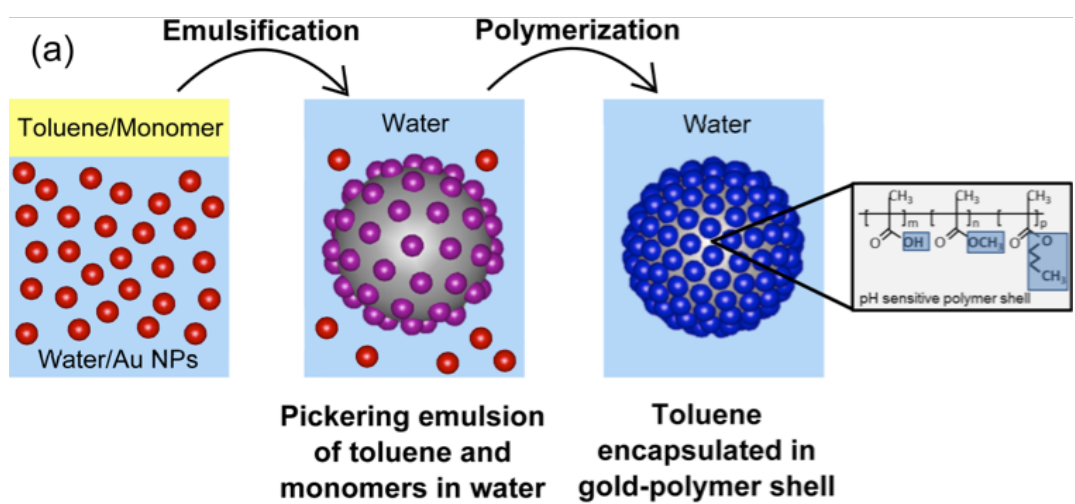

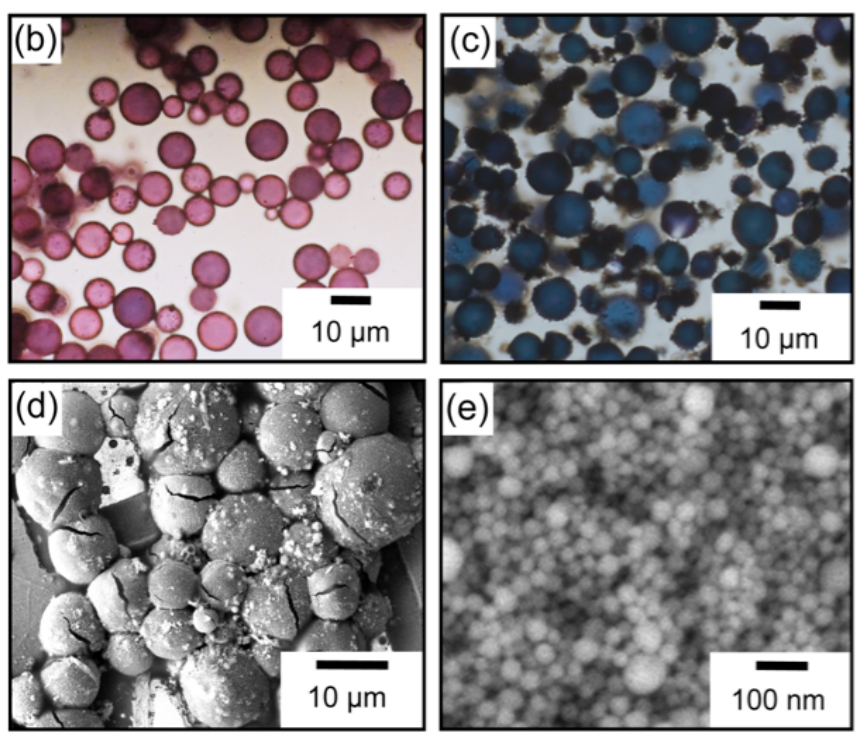

**Figure 2**: (a) Schematic representation of the formation of plasmocapsules. (b) Optical microscope images of typical emulsions right after sonication. (c) Optical microscope images of typical plasmocapsules obtained after polymerization of the emulsion. (d) SEM images of dried plasmocapsules. (e) SEM image of densely packed Au NPs embedded in the polymer shell of the plasmocapsules.

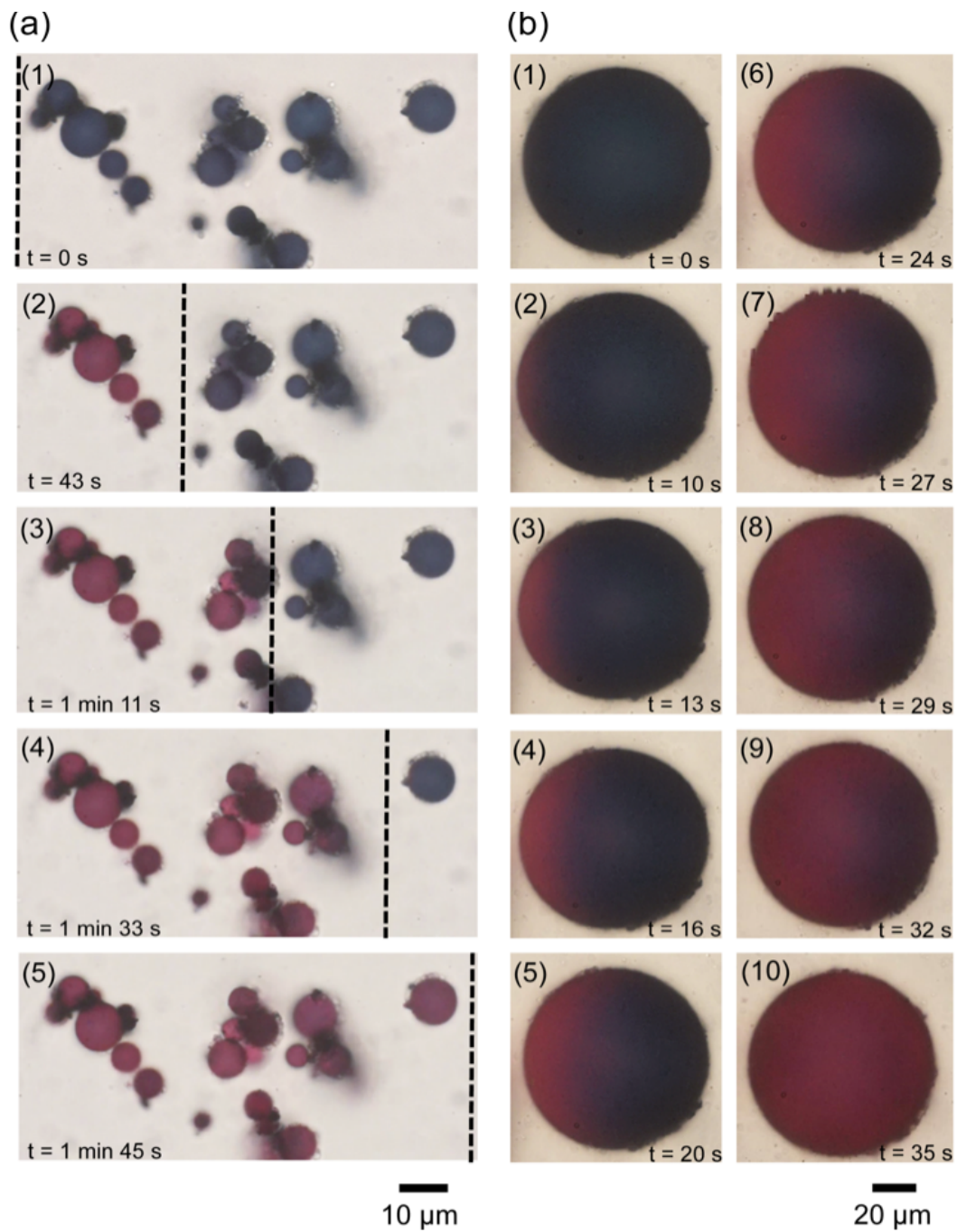

**Figure 3**: (a) Plasmocapsules swelling. Initially in acidic solution (pH=1). The pH is slowly increased by flowing a solution of sodium hydroxide (pH=13.8, $10^{-1}$ M) from left to right (1-5); the dashed black line represents the progression of the pH front. (b) Color change of an individual plasmocapsule-swelling as a function of increasing pH.

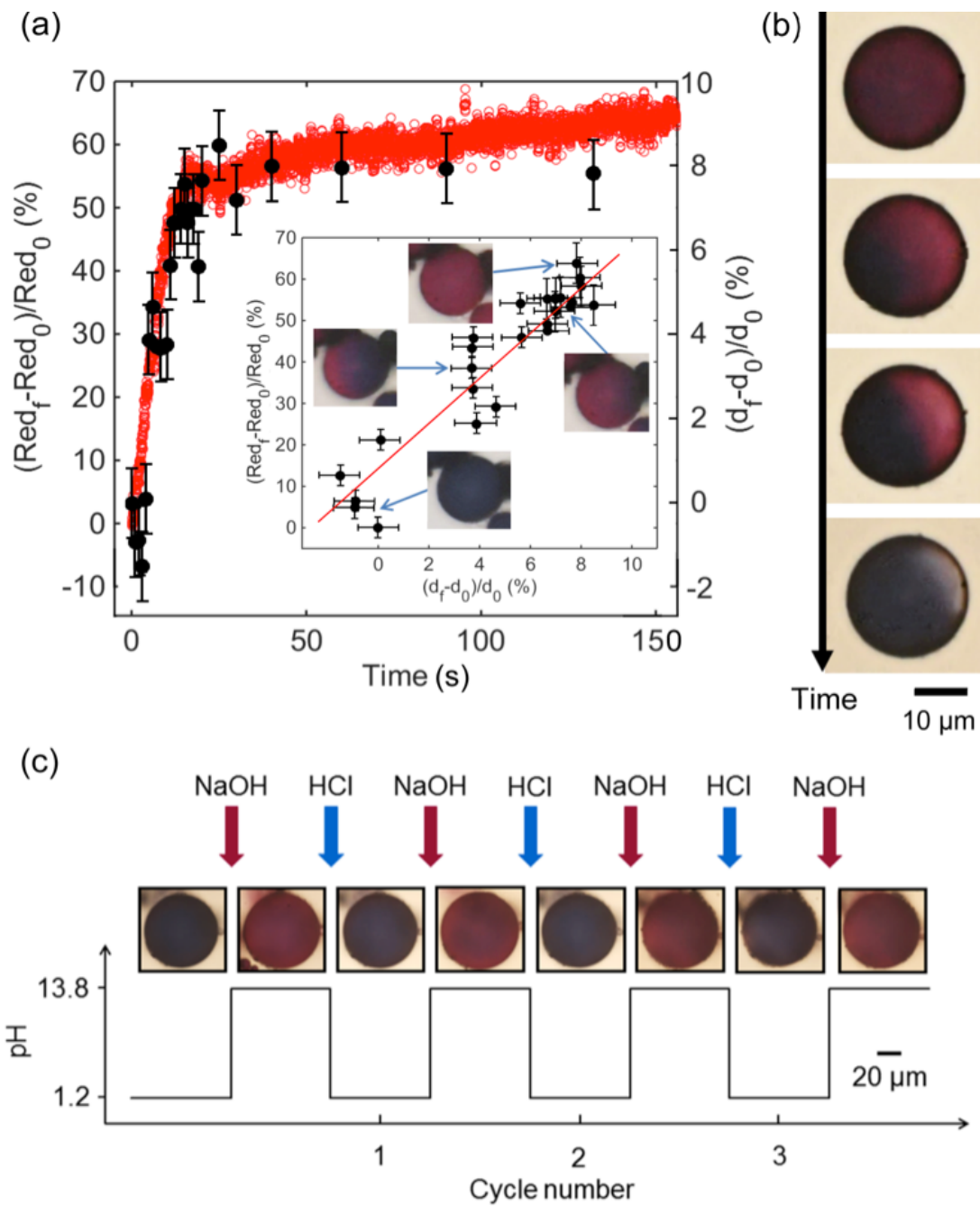

**Figure 4**: (a) Plasmocapsules swelling and change of color over time (microcapsule diameter is 12 μm). Insert: Plasmocapsule change of color as a function of its increase in diameter. (b) Reversibility of swelling and color change upon pH decrease. (c) Plasmocapsule color change with pH cycles.

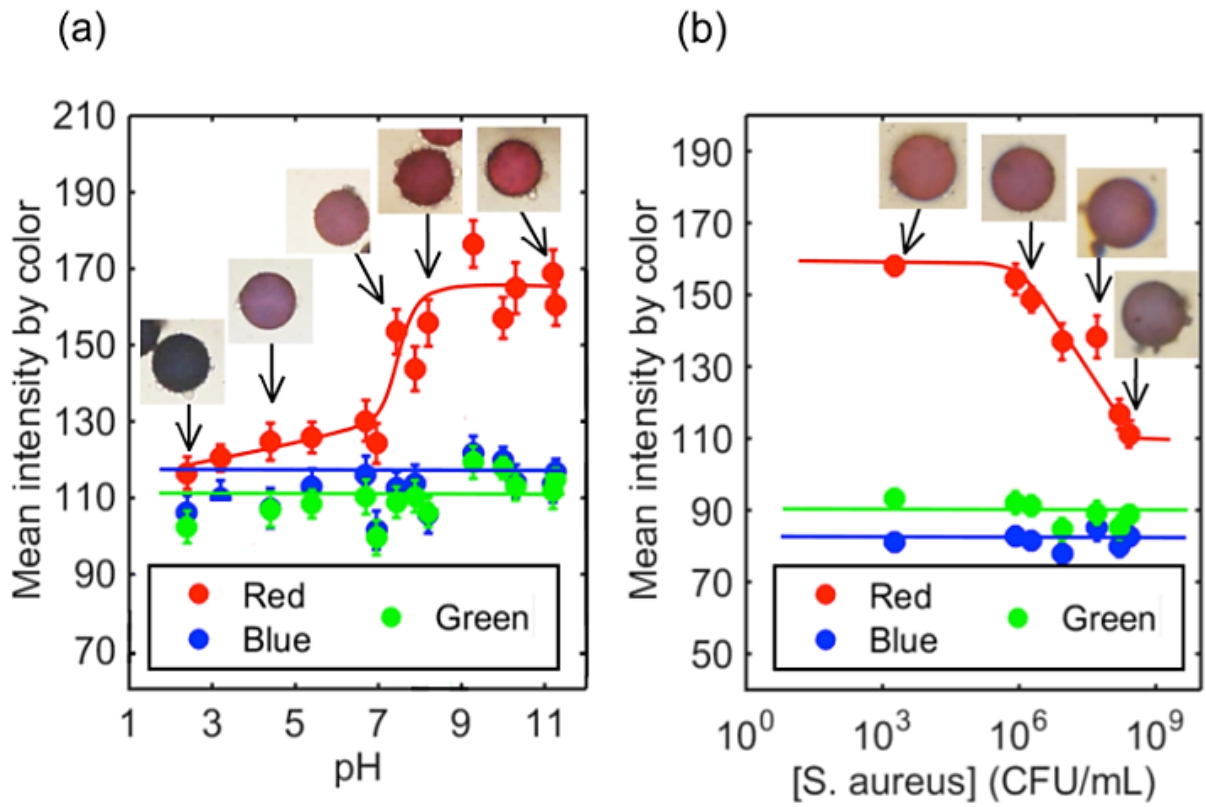

**Figure 5**: (a) Change of color of the plasmocapsules as a function of (a) the pH and (b) the concentration of S. aureus.

**ToC**:

Plasmocapsules made of optically active plasmonic gold nanoparticles and pH-responsive polyacrylate are used as pH microsensors. Upon pH change, the polymer shell of the plasmocapsules swells or shrinks. Concomitantly, the distance between the gold nanoparticles embedded in the polymeric matrix varies, resulting in a color change. Each single plasmocapsule is a reversible independent microsensor over a large range of pH.

**Keyword:** Microsensors

C.A. S. Burel[1], A.Teolis[1], A. Alsayed[1], B. Donnio[1,2], and R. Dreyfus[1]*

**Dispersible Elastic Plasmocapsules as Reversible pH-microsensor Based on Color Change**

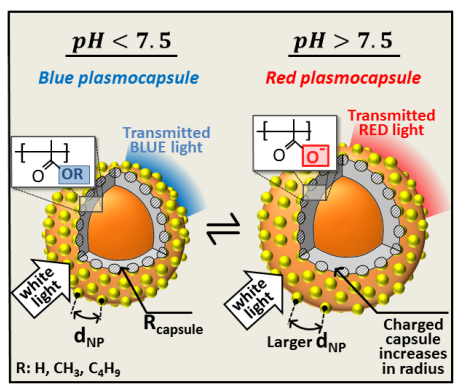